\documentclass[preprint,12pt]{elsarticle}

\usepackage{amssymb}
\usepackage{multicol}
\journal{Nucl. Inst. and Meth. A Proceedings}

\begin{document}

\begin{frontmatter}



\title{Development of Hybrid Resistive Plate Chambers}


\author[inst1,inst2,inst3]{M. Tosun}
\author[inst1,inst2,inst4]{B. Bilki}
\author[inst1,inst2,inst5]{K. K. Sahbaz}

\affiliation[inst1]{organization={Beykent University, Istanbul}
,%Department and Organization
            country={Turkey}
}
\affiliation[inst2]{organization={Turkish Accelerator and Radiation Laboratory, Ankara},%Department and Organization
            country={Turkey}
}
\affiliation[inst3]{organization={Yildiz Technical University, Istanbul},%Department and Organization
            country={Turkey}
}
\affiliation[inst4]{organization={University of Iowa, Iowa City, IA},%Department and Organization
            country={USA}
}
\affiliation[inst5]{organization={Ankara University, Ankara},%Department and Organization
            country={Turkey}
}

\begin{abstract}
Resistive Plate Chambers (RPCs) are essential active media of large-scale experiments as part of the muon systems and (semi-)digital hadron calorimeters. Among the several outstanding issues associated with the RPCs, the loss of efficiency for the detection of particles when subjected to high particle fluxes, and the limitations associated with the common RPC gases can be listed. In order to address the latter issue, we developed novel RPC designs with special anode plates coated with high secondary electron emission yield materials such as Al$_2$O$_3$ and TiO$_2$. The proof of principle was obtained for various designs and is in progress for the rest. The idea was initiated following the achievements on the development of the novel 1-glass RPCs.

Here we report on the construction of various different RPC designs, and their performance measurements in laboratory tests and with particle beams; and discuss the future test plans.
\end{abstract}




\end{frontmatter}


\section{Introduction}
\label{sec:introduction}

Resistive Plate Chambers (RPCs) were introduced in the 1980s \cite{ref-santonico} and have been widely used by the High Energy Physics community since then. The RPCs have been utilized in large-scale experiments mostly as triggering and precision timing detectors. The working principle of the RPCs relies on the ionization in a thin gas gap which is provided by two or more resistive plates of high resistivity like glass or Bakelite. The RPCs are provided with a high voltage on the outer surfaces so that the primary ionization is multiplied in the gas gap. The signal is picked up by either strips or pads, which are placed on the outside of the chambers. The RPCs also utilized as the active media of the CALICE \cite{ref-calice} (semi-)digital hadron calorimeters \cite{ref-dhcalconstruction,ref-sdhcal}.

In novel designs of RPCs, only one resistive plate is used and the signal pickup is performed inside the chamber. Several 1-glass RPCs were built and tested to date, and their performance as single particle detectors as well as calorimeter active media was validated in beam tests \cite{ref-1g}. The placement of the anode plane inside the chamber enables the possibility of exploring functional anodes by applying surface coatings on the pickup pads or strips. Given the limitations in the utilization of the common RPC gases such as r134a and SF$_6$ due to their high global warming potential, this functionality can be utilized to reduce or completely abandon the flow of the common RPC gases, or to introduce alternative gases with negligible greenhouse effects. In order to explore this implementation, we developed the hybrid RPCs where part of the electron multiplication is transferred from the gas layer to a solid state layer coated on the surface of the anode as a thin film. The thin film materials are chosen to have high secondary electron multiplication yields, such as Al$_2$O$_3$ and TiO$_2$.

Here we describe the details of the first generation hybrid RPCs and their performance measurements in laboratory tests and with particle beams; and provide a perspective for future directions.

\section{Construction of the Hybrid RPCs}
\label{sec:construction}

\begin{figure}
\centering
\includegraphics[height=5 cm]{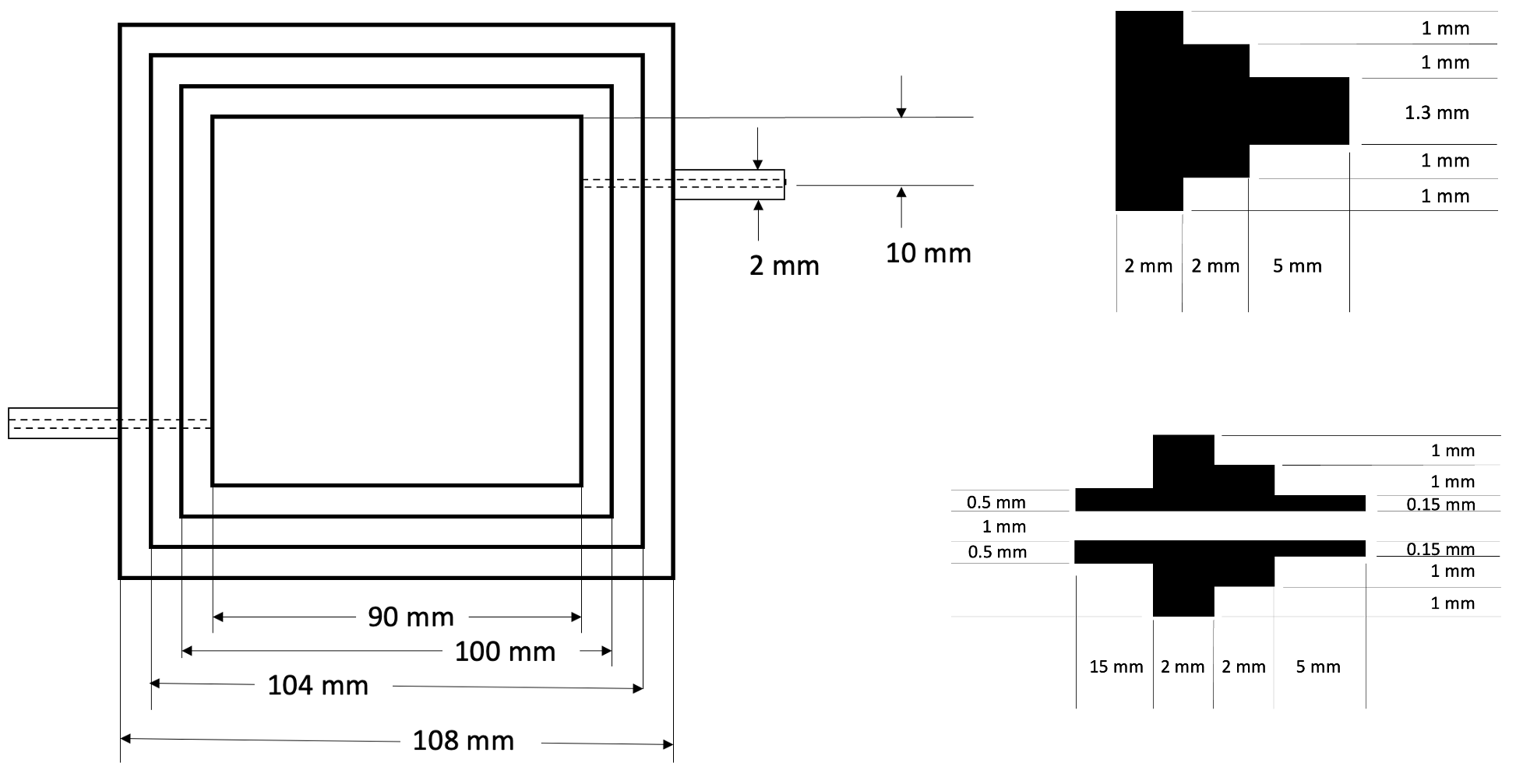}
\caption{The sketches of the RPC frame (left) and the cross sections of a regular section (top right) and a gas inlet/outlet section (bottom right). Not to scale. \label{fig_1}}
\end{figure}

Several 10 cm $\times$ 10 cm chambers were made with 2 mm thick glass plates and a single readout pad of size 9 cm $\times$ 9 cm. The 5 mm rim of the glass plates were masked and a mixture of a high resistivity and a low resistivity artist paint was applied with an airbrush gun to yield 1 - 5 M$\Omega/\square$ surface resistivity. The RPC frame was 3D-printed and the glass plates and the pad boards were glued with two-component epoxy. The gas gap was 1.3 mm. Figure \ref{fig_1} shows the sketches of the RPC frame (left) and the cross sections of a regular section (top right) and a gas inlet/outlet section (bottom right).

\begin{figure}
\centering
\includegraphics[height=3.5 cm]{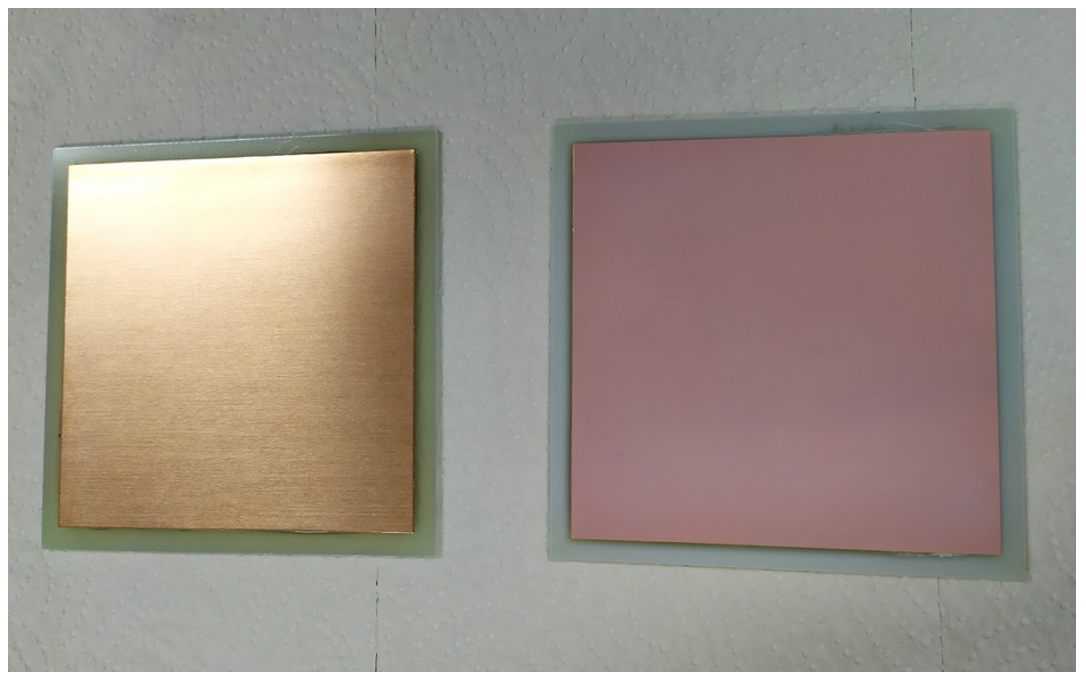}
\caption{A picture of the clean (left) and the TiO$_2$-coated (right) RPC pad boards. \label{fig_2}}
\end{figure}

\begin{figure}
\centering
\includegraphics[height=3.5 cm]{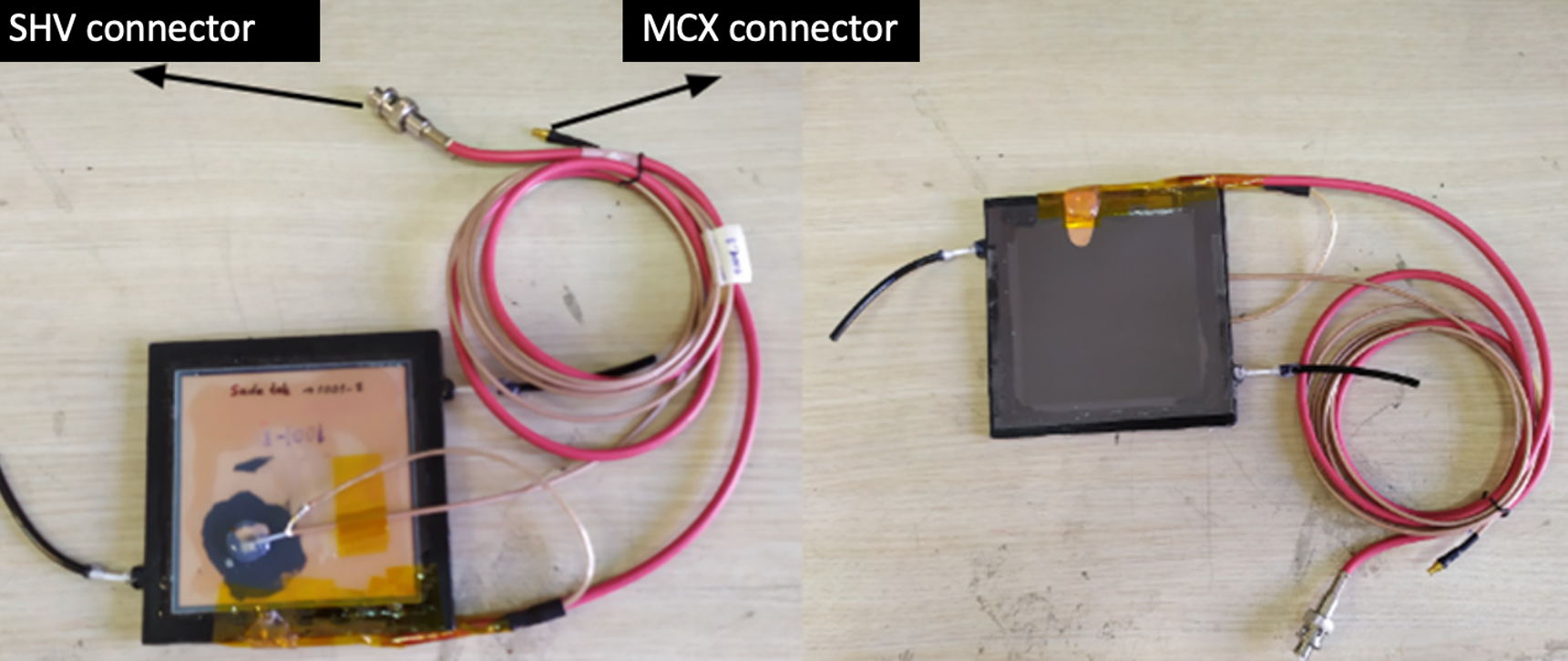}
\caption{The pictures of the front (right) and back (left) sides of fully assembled RPCs. \label{fig_3}}
\end{figure}

The coating of Al$_2$O$_3$ on the pad boards was done with magnetron sputtering at Gazi University Photonics Application and Research Center, Ankara, Turkey \cite{ref-gazi}. Two different thicknesses were applied, 500 nm and 350 nm. The RPCs constructed with these pad boards were labeled as Al$_2$O$_3$-v1 and Al$_2$O$_3$-v2 respectively. The coating of TiO$_2$ was made in the laboratory by dissolving the TiO$_2$ powder in ethanol and applying the solution on the pad board with an airbrush. The physical properties of the solution are closer to a fine dispersion and a uniform coating with sufficiently good adhesion can be obtained on the pad boards with the airbrush. Three different surface densities were implemented; 1 mg/cm$^2$, 0.5 mg/cm$^2$ and 0.15 mg/cm$^2$; with the RPCs constructed with these pad boards labelled as TiO$_2$-v1, TiO$_2$-v2 and TiO$_2$-v3 respectively. Figure \ref{fig_2} shows a picture of the clean (left) and the TiO$_2$-coated (right) RPC pad boards. Figure \ref{fig_3} shows the pictures of the front (right) and back (left) sides of fully assembled RPCs. The RPCs were then subjected to gas leak and HV tests. The operating performance was validated in a cosmic muon test station for an extended period.

\section{Beam Tests of the Hybrid RPCs}
\label{sec:beamtests}

The hybrid RPCs and one standard 1-glass and one standard 2-glass RPC were exposed to the broadband muon beam of Fermilab Test Beam Facility \cite{ref-ftbf}. The gas mixture was r134a:isobutane:SF$_6$ 94.5:5.0:0.5 with a flow rate of 2-3 cc/min, approximately half the common value of 5 cc/min.

\begin{figure}
    \centering
    \begin{minipage}{0.45\textwidth}
        \centering
        \includegraphics[width=0.9\textwidth]{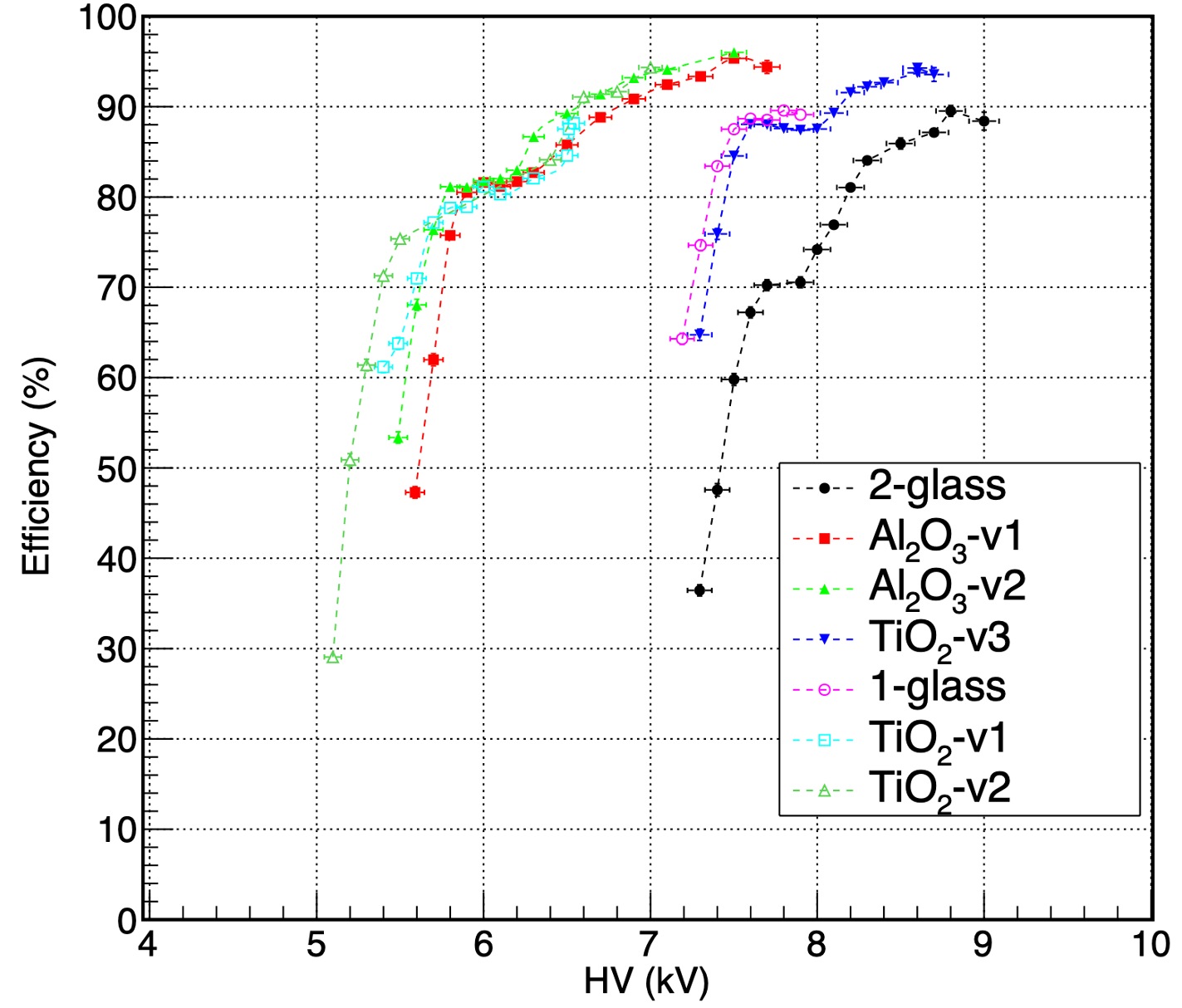}        \caption{The efficiency as a function of the applied high voltage for the seven types of RPCs tested (see text). \label{fig_4}}
    \end{minipage}\hfill
    \begin{minipage}{0.45\textwidth}
        \centering
        \includegraphics[width=0.9\textwidth]{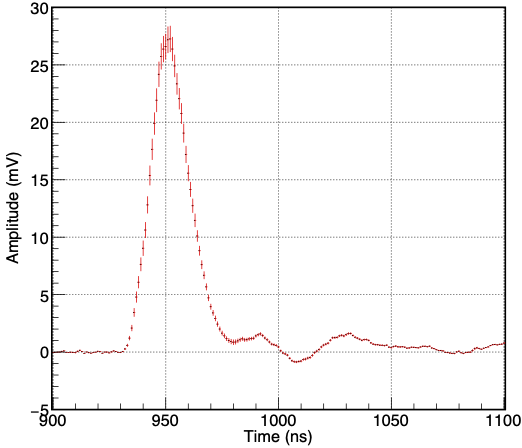}
        \caption{The average waveform of the Al$_2$O$_3$-v1 at 7.3 kV for the MIP events when the RPC is efficient. \label{fig_5}}
    \end{minipage}
\end{figure}

Figure \ref{fig_4} shows the efficiency as a function of the applied high voltage for the seven types of RPCs tested. The efficiency threshold was set at 300 fC. If one considers the 90 \% efficiency crossing high voltage value as a measure, the highest value is for the standard 2-glass RPC around 8.5 kV. The standard 1-glass RPC and the TiO$_2$-v3, which has the lowest surface density of TiO$_2$ have comparable values around 7.5 kV. The major impact is observed for the RPCs with higher surface density TiO$_2$ coated and Al$_2$O$_3$ coated anodes. The 90 \% efficiency crossing voltage for these hybrid RPCs is around 6.5 kV with RPCs still retaining higher than 50 \% efficiency down to around 5 kV. There is a clear indication of the contribution of the electron multiplication in the applied coating. The electron multiplication in the secondary emission layer is prompt. Figure \ref{fig_5} shows the average waveform of the Al$_2$O$_3$-v1 at 7.3 kV for the MIP events when the RPC is efficient. The average waveform does not exhibit any distortion of the RPC pulse shape. 

\section{Conclusions}
\label{sec:conclusions}
The 1-glass RPC design offers several advantages over 2-glass RPCs such as avalanches with higher collimation and an increased rate capability. It also starts a new chapter where the in-chamber anode plate can be made more functional. By coating the anode plate with high secondary electron yield materials, electron multiplication in the chamber can be enhanced considerably. R\&D is underway to fully characterize the newly developed, so-called hybrid RPCs. The hybrid RPCs have the potential to mitigate the limitations associated with the RPC gases and to relax the overall operating conditions. Long-term stability tests, timing measurements and the measurements of the response with alternative gas mixtures are underway.

\section*{Acknowledgements}
This work is supported by Tübitak Grant No: 118C224. 

%\appendix

%\section{Sample Appendix Section}
%Lorem ipsum dolor sit amet, consectetur adipiscing elit, sed do eiusmod tempor section \ref{sec:sample1} incididunt ut labore et dolore magna aliqua. Ut enim ad minim veniam, quis nostrud exercitation ullamco laboris nisi ut aliquip ex ea commodo consequat. Duis aute irure dolor in reprehenderit in voluptate velit esse cillum dolore eu fugiat nulla pariatur. Excepteur sint occaecat cupidatat non proident, sunt in culpa qui officia deserunt mollit anim id est laborum.

% \bibliographystyle{elsarticle-num} 
% \bibliography{cas-refs}


\end{document}